\RequirePackage{lineno}
\documentclass[aps,prd,twocolumn,showpacs,superscriptaddress]{revtex4-1}

\usepackage{amssymb}
\usepackage{amsmath}

\usepackage{graphicx}
\usepackage{dcolumn}
\usepackage{bm}
\usepackage{pstricks,pst-node,pst-text,pst-3d}

\usepackage{psfrag}
\usepackage{xspace}

\usepackage{natbib}
\usepackage{url}

\usepackage{ifpdf}
\ifpdf
\DeclareGraphicsExtensions{.pdf, .jpg, .tif}
\usepackage[%
  pdftitle={Reevaluation of the parton distribution of strange quarks in the nucleon},%
  pdfauthor={The HERMES Collaboration},%
  pdfsubject={HERMES Strangeness},%
  pdfstartview=FitH,%
  bookmarks=true,%
  bookmarksopen=true,%
  breaklinks=true,%
  colorlinks=true,%
  linkcolor=blue,anchorcolor=blue,%
  citecolor=blue,filecolor=blue,%
  menucolor=blue,pagecolor=blue,%
  urlcolor=blue]{hyperref}
\else
\DeclareGraphicsExtensions{.eps, .jpg}
\usepackage[%
  breaklinks=true,%
  colorlinks=true,%
  linkcolor=blue,anchorcolor=blue,%
  citecolor=blue,filecolor=blue,%
  menucolor=blue,pagecolor=blue,%
  urlcolor=blue]{hyperref}
\fi

\newcommand{\pt}{\ensuremath{P_{h\perp}}\xspace}

\newcommand{\ubar}{\ensuremath{\bar{u}}\xspace}
\newcommand{\dbar}{\ensuremath{\bar{d}}\xspace}
\newcommand{\sbar}{\ensuremath{\bar{s}}\xspace}

\begin{document}


\title{Reevaluation of the parton distribution of strange quarks in the nucleon}


\def\groupargonne{\affiliation{Physics Division, Argonne National Laboratory, 
Argonne, Illinois 60439-4843, USA}}
\def\groupbari{\affiliation{Istituto Nazionale di Fisica Nucleare, 
Sezione di Bari, 70124 Bari, Italy}}
\def\groupbeijing{\affiliation{School of Physics, Peking University, Beijing 
100871, China}}
\def\groupbilbao{\affiliation{Department of Theoretical Physics, University of 
the Basque Country UPV/EHU, 48080 Bilbao, Spain and IKERBASQUE, 
Basque Foundation for Science, 48011 Bilbao, Spain}}
\def\groupcolorado{\affiliation{Nuclear Physics Laboratory, University of 
Colorado, Boulder, Colorado 80309-0390, USA}}
\def\groupdesy{\affiliation{DESY, 22603 Hamburg, Germany}}
\def\groupzeuthen{\affiliation{DESY, 15738 Zeuthen, Germany}}
\def\groupdubna{\affiliation{Joint Institute for Nuclear Research, 141980 Dubna,
 Russia}}
\def\grouperlangen{\affiliation{Physikalisches Institut, Universit\"at 
Erlangen-N\"urnberg, 91058 Erlangen, Germany}}
\def\groupferrara{\affiliation{Istituto Nazionale di Fisica Nucleare, 
Sezione di Ferrara and Dipartimento di Fisica e Scienze della Terra, 
Universit\`a di Ferrara, 44122 Ferrara, Italy}}
\def\groupfrascati{\affiliation{Istituto Nazionale di Fisica Nucleare, 
Laboratori Nazionali di Frascati, 00044 Frascati, Italy}}
\def\groupgent{\affiliation{Department of Physics and Astronomy, Ghent 
University, 9000 Gent, Belgium}}
\def\groupgiessen{\affiliation{II. Physikalisches Institut, 
Justus-Liebig Universit\"at Gie{\ss}en, 35392 Gie{\ss}en, Germany}}
\def\groupglasgow{\affiliation{SUPA, School of Physics and Astronomy, University
 of Glasgow, Glasgow G12 8QQ, United Kingdom}}
\def\groupillinois{\affiliation{Department of Physics, University of Illinois, 
Urbana, Illinois 61801-3080, USA}}
\def\groupmichigan{\affiliation{Randall Laboratory of Physics, University of 
Michigan, Ann Arbor, Michigan 48109-1040, USA }}
\def\groupmoscow{\affiliation{Lebedev Physical Institute, 117924 Moscow, Russia}
}
\def\groupnikhef{\affiliation{National Institute for Subatomic Physics (Nikhef),
 1009 DB Amsterdam, The Netherlands}}
\def\groupstpetersburg{\affiliation{B.P. Konstantinov Petersburg Nuclear 
Physics Institute, Gatchina, 188300 Leningrad region, Russia}}
\def\groupprotvino{\affiliation{Institute for High Energy Physics, Protvino, 
142281 Moscow region, Russia}}
\def\groupregensburg{\affiliation{Institut f\"ur Theoretische Physik, 
Universit\"at Regensburg, 93040 Regensburg, Germany}}
\def\grouprome{\affiliation{Istituto Nazionale di Fisica Nucleare, Sezione di Roma, Gruppo Collegato Sanit\`a and Istituto Superiore di Sanit\`a, 00161 Roma, 
Italy}}
\def\grouptriumf{\affiliation{TRIUMF, Vancouver, British Columbia V6T 2A3, 
Canada}}
\def\grouptokyo{\affiliation{Department of Physics, Tokyo Institute of 
Technology, Tokyo 152, Japan}}
\def\groupamsterdam{\affiliation{Department of Physics and Astronomy, VU 
University, 1081 HV Amsterdam, The Netherlands}}
\def\groupwarsaw{\affiliation{National Centre for Nuclear Research, 00-689 
Warsaw, Poland}}
\def\groupyerevan{\affiliation{Yerevan Physics Institute, 375036 Yerevan, 
Armenia}}
\def\groupnone{\noaffiliation}


\groupargonne
\groupbari
\groupbeijing
\groupbilbao
\groupcolorado
\groupdesy
\groupzeuthen
\groupdubna
\grouperlangen
\groupferrara
\groupfrascati
\groupgent
\groupgiessen
\groupglasgow
\groupillinois
\groupmichigan
\groupmoscow
\groupnikhef
\groupstpetersburg
\groupprotvino
\groupregensburg
\grouprome
\grouptriumf
\grouptokyo
\groupamsterdam
\groupwarsaw
\groupyerevan


\author{A.~Airapetian}  \groupgiessen \groupmichigan
\author{N.~Akopov}  \groupyerevan
\author{Z.~Akopov}  \groupdesy
\author{E.C.~Aschenauer}\thanks{Now at: Brookhaven National Laboratory, Upton, New York 11772-5000, USA}  \groupzeuthen
\author{W.~Augustyniak}  \groupwarsaw
\author{A.~Avetissian}  \groupyerevan
\author{E.~Avetisyan}  \groupdesy
\author{S.~Belostotski}  \groupstpetersburg
\author{H.P.~Blok}  \groupnikhef \groupamsterdam
\author{A.~Borissov}  \groupdesy
% ...

\collaboration{HERMES Collaboration}

\date{\today}


\begin{abstract}

\newpage
An earlier extraction from the HERMES experiment
of the polarization-averaged 
parton distribution of strange quarks
in the nucleon has been reevaluated  using final data on the multiplicities 
of charged kaons in semi-inclusive deep-inelastic 
scattering obtained with a kinematically more comprehensive
method of correcting for experimental effects. 
General features of the distribution are
confirmed, but the rise at low $x$ is less pronounced than previously
reported.

\begin{description}
\item[PACS numbers:]
13.60.-r, 13.88.+e, 14.20.Dh, 14.65.-q
\item[Keywords:]
Parton, strange quark, parton distribution
\end{description}
\end{abstract}

\maketitle

The parton distribution functions (PDFs) of the strange quarks in the
nucleon describe important features of the structure of 
the quark sea, and constrain models of its 
origin~\cite{Bjorken:1968dy,Feynman:1973xc,Gluck:1988xx,Gluck:1991ng}. 
In addition, the strangeness content of the nucleon is of interest
because of its impact on calculations of short-distance processes at
high energies~\cite{Kusina:2012vh}, {and also 
in view of recent ATLAS results~\cite{Aad:2012sb}, 
which suggest that at small $x$ it could be substantially 
larger than previously assumed.} 
In 2008 HERMES 
published the results of the extraction
of the momentum and helicity density distributions 
of the strange sea in the nucleon from charged-kaon production in
deep-inelastic scattering (DIS) on the deuteron~\cite{Airapetian:2008qf}.
The shape of the polarization-averaged distribution in $x$,
where $x$ is the dimensionless
Bjorken scaling variable,
was observed to be softer than that of the average of
the \ubar and \dbar quarks.
The helicity distribution was found to be compatible with
zero in the region of 
measurement 0.02 $<$ $x$ $<$ 0.60.

HERMES has finalized the extraction of multiplicities for each charged
state of $\pi^{\pm}$ and K${^{\pm}}$~\cite{Airapetian:2012ki}. 
In the extraction, the correction
for acceptance, kinematic smearing, losses due to decay in flight 
and secondary strong interactions, and radiative effects is accomplished
by means of a smearing matrix which is generated with a Monte Carlo
simulation. The procedure is described in detail 
in Ref.~\cite{Airapetian:2012ki}. The data of
Ref.~\cite{Airapetian:2008qf} were obtained 
by carrying out the unfolding to correct for these 
effects in only one dimension, $x$. Further study of the unfolding
procedure has revealed that using a multi-dimensional unfolding in
$x$, $z$, and \pt results in significant 
changes in the final multiplicities.
Here $z\equiv E_h/\nu$ with $\nu$ and $E_h$ the energies  of the virtual 
photon and of the detected hadron in the target rest frame, respectively,
and $P_{h\perp}$ is the transverse momentum of the hadron with respect
to the virtual-photon direction.   
The results for the final multiplicities~\cite{Airapetian:2012ki}
were obtained with this improvement and with
the elimination of a requirement, used in earlier extractions, that
the hadrons have momenta greater than 2 GeV. {
In practice, multiplicities are not defined with such a
limit on the integration over the hadron momentum.}  
The purpose of this brief note is to update the 
extraction of the strange-quark PDF
$S(x)\equiv s(x) + \sbar(x)$ 
by employing the final HERMES multiplicities obtained by this advanced
analysis. The extraction is carried out  
in leading logarithmic order (LO) in the strong coupling constant of
quantum chromodynamics.  
While a next-to-leading order (NLO) extraction
would be preferred, such a procedure using semi-inclusive DIS data
is not currently available. However, because of the wide interest in
shape and magnitude of $S(x)$, a LO extraction is an important first step.

\begin{figure}[b]
	\centering
	\includegraphics[width=0.89\linewidth]{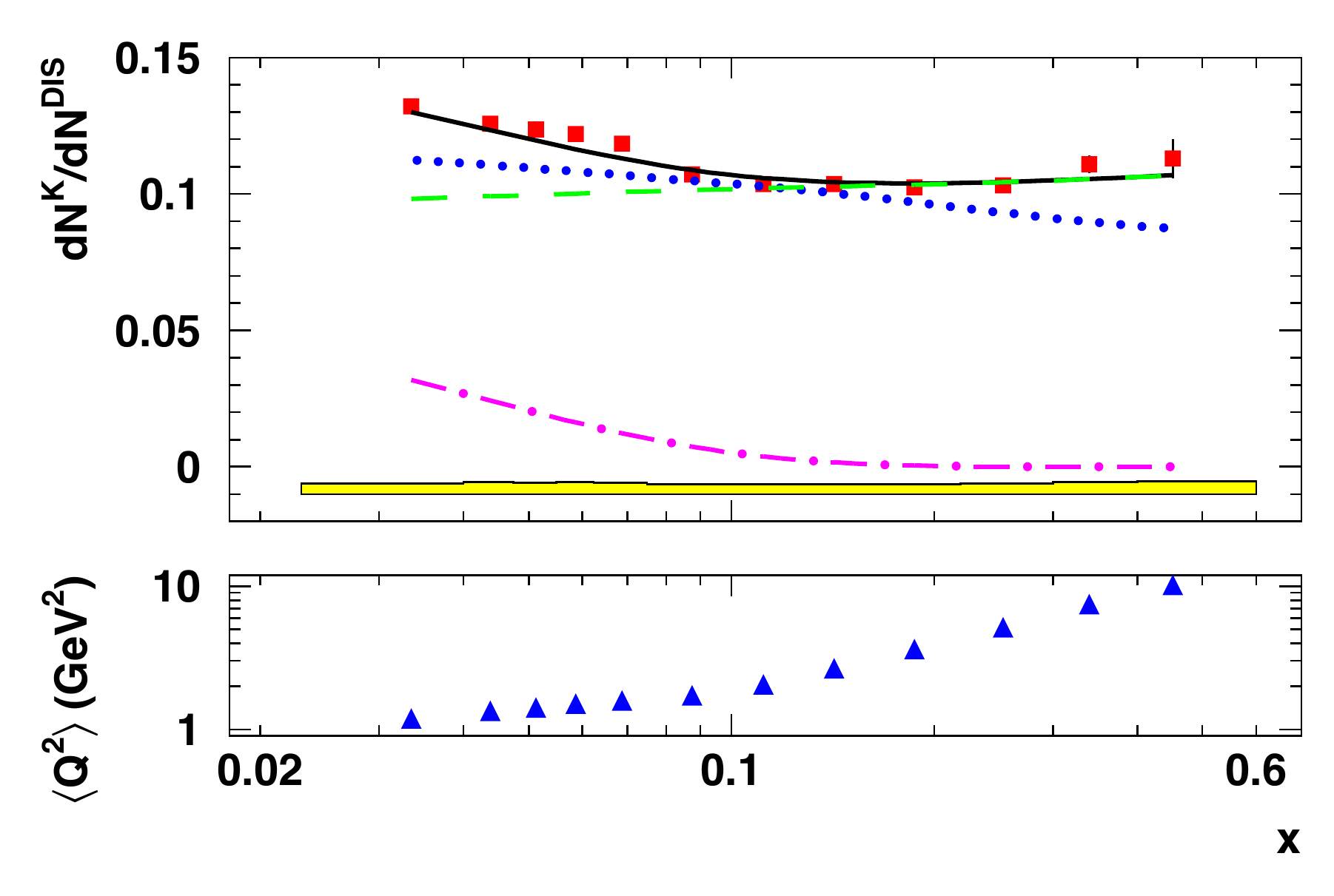}
	\caption{The multiplicity of charged kaons
         in semi-inclusive DIS from a deuterium target, as a function
         of Bjorken $x$. The continuous curve is calculated from 
         the strange-quark contribution taken from the fit  
         in Fig.~\ref{fig:sxdz}, together with 
         the nonstrange contribution as extracted
         from the high-$x$ multiplicity data (see text).
         The green dashed (magenta dash-dotted) curve shows separately
         the nonstrange- (strange-) 
         quark contribution to the multiplicity for that fit.
         The blue-dotted curve is the LO prediction obtained with 
         {CTEQ6L} PDFs and fragmentation functions 
         from~\cite{deFlorian:2007aj}.
         The values of $\langle {\mathrm{Q}}^2 \rangle$ for
         each $x$ bin are shown in the lower panel.
         The band represents systematic uncertainties.}
	\label{fig:mult_x}
\end{figure}

In the isoscalar method used in the HERMES measurement, 
the distribution in $x$ 
of the strange-quark sea is extracted from the spin-averaged kaon,
$K \equiv K^++K^-$, multiplicity for the DIS of positrons/electrons by a
deuteron target~\cite{Airapetian:2008qf}. 
For the isoscalar deuteron, in LO this observable
depends on the PDFs $S(x)$ and 
$Q(x)\equiv u(x)+\ubar (x)+d(x)+\dbar (x)$.
Technical details of the experiment and the principles of the 
procedure for extracting the density distributions for these quantities
are presented in Ref.~\cite{Airapetian:2008qf}. The extracted kaon 
multiplicity is shown in Fig.~\ref{fig:mult_x} as a function of $x$
at the corresponding average ${\mathrm{Q}}^2$ of each bin. 

The multiplicity presented in Fig.~\ref{fig:mult_x} is the
starting point for the extraction of $S(x,\mathrm{Q}^2)$, and can be used in
the future, e.g., for NLO analyses. 
Here we update the extraction of $S(x,\mathrm{Q}^2)$ 
which is based on a number
of simplifying assumptions, i.e., LO and  leading twist, as well as on fixing
the fragmentation functions and PDFs to a specific set.
As in the analysis reported earlier, both $S(x,{\mathrm{Q}}^2)$ and the quantity
$\int {\cal D}_S^{K}(z,{\mathrm{Q}}^2)dz$ are taken as unknown, and the analysis
is carried out extracting their product.
Here, $\int {\cal D}_S^{K}(z,{\mathrm{Q}}^2)dz$ is the integral over the measured
region of $z$ of the fragmentation
function describing the number density of charged kaons from a 
struck quark of flavor $S$, with
${\cal D}^K_{S}(z) \equiv  2{{\cal D}}_s^{K}(z)$ by charge conjugation
symmetry.
The relationship between this product and the kaon multiplicity
is given in LO by~\cite{Airapetian:2008qf}
\begin{eqnarray}\label{eq:mult}
\lefteqn{\frac{dN^{K}(x,\mathrm{Q}^2)}{dN^{DIS}(x,\mathrm{Q}^2)} = } \\
& & \frac{Q(x,\mathrm{Q}^2){\int{\cal D}^{K}_{Q}(z,{\mathrm{Q}}^2)dz}+
 S(x,{\mathrm{Q}}^2){\int{\cal D}^{K}_{S}(z,{\mathrm{Q}}^2)dz} }  
{5Q(x,\mathrm{Q}^2)+2S(x,{\mathrm{Q}}^2)}. \nonumber
\end{eqnarray}
An absence of $x$ dependence shown by the HERMES data above $x>0.1$ 
requires that $d[S(x))/Q(x)]/dx$ vanish. This indicates that either
$S(x)=0$ or $S(x)=kQ(x)$ where $k$ is a constant. $S(x)$ is not expected
to have the same shape as $Q(x)$, and all PDF reference data sets published
to date are consistent with this expectation. Therefore for $x>0.1$ it is
concluded that $S(x,{\mathrm{Q}}^2)$$\approx$$0$.

In the limit $S(x,{\mathrm{Q}}^2)$$\rightarrow$$0$, 
for a deuteron target the multiplicity
${dN^{K}(x,{\mathrm{Q}}^2)}/{dN^{DIS}(x,{\mathrm{Q}}^2)} = 
\int{\cal D}^{K}_{Q}(z,\mathrm{Q}^2)dz/5$ 
(see Eq.~\ref{eq:mult}).
For $x>0.1$ the multiplicity measured by HERMES 
is almost constant at a value of about 0.1. 
The value of the multiplicity throughout this region provides a direct
estimate of $\int{\cal D}^{K}_{Q}(z,\mathrm{Q}^2)dz$.
To account for any
residual dependence on ${\mathrm{Q}}^2$ or, because of the correlation 
between $x$ and Q$^2$, equivalently 
on $x$, a first-degree 
polynomial was fitted to the multiplicity for $x$$>$$0.1$ yielding the 
result that $dN^K(x,{\mathrm{Q}}^2)/dN^{DIS}(x,{\mathrm{Q}}^2)$ = 
(0.102$\pm$0.002)+(0.013$\pm$0.010)$x$, 
as shown by the green dashed curve in Fig.~\ref{fig:mult_x}.
In the region near $x$=0.13, where ${\mathrm{Q}}^2\approx 2.5$ GeV$^2$,
this fit gives the result 
$\int_{0.2}^{0.8}{\cal D}^K_Q(z,{\mathrm{Q}}^2)dz=0.514 \pm 0.010$, 
compared to the value $0.435\pm 0.044$
obtained for ${\mathrm{Q}}^2=2.5$ GeV$^2$ from a
global analysis of fragmentation functions~\cite{deFlorian:2007aj}.
The weak $x$ dependence obtained in the fit is 
consistent with the ${\mathrm{Q}}^2$
dependence exhibited by the results of the global analysis.

\begin{figure}[h]
	\centering
	\includegraphics[width=0.89\linewidth]{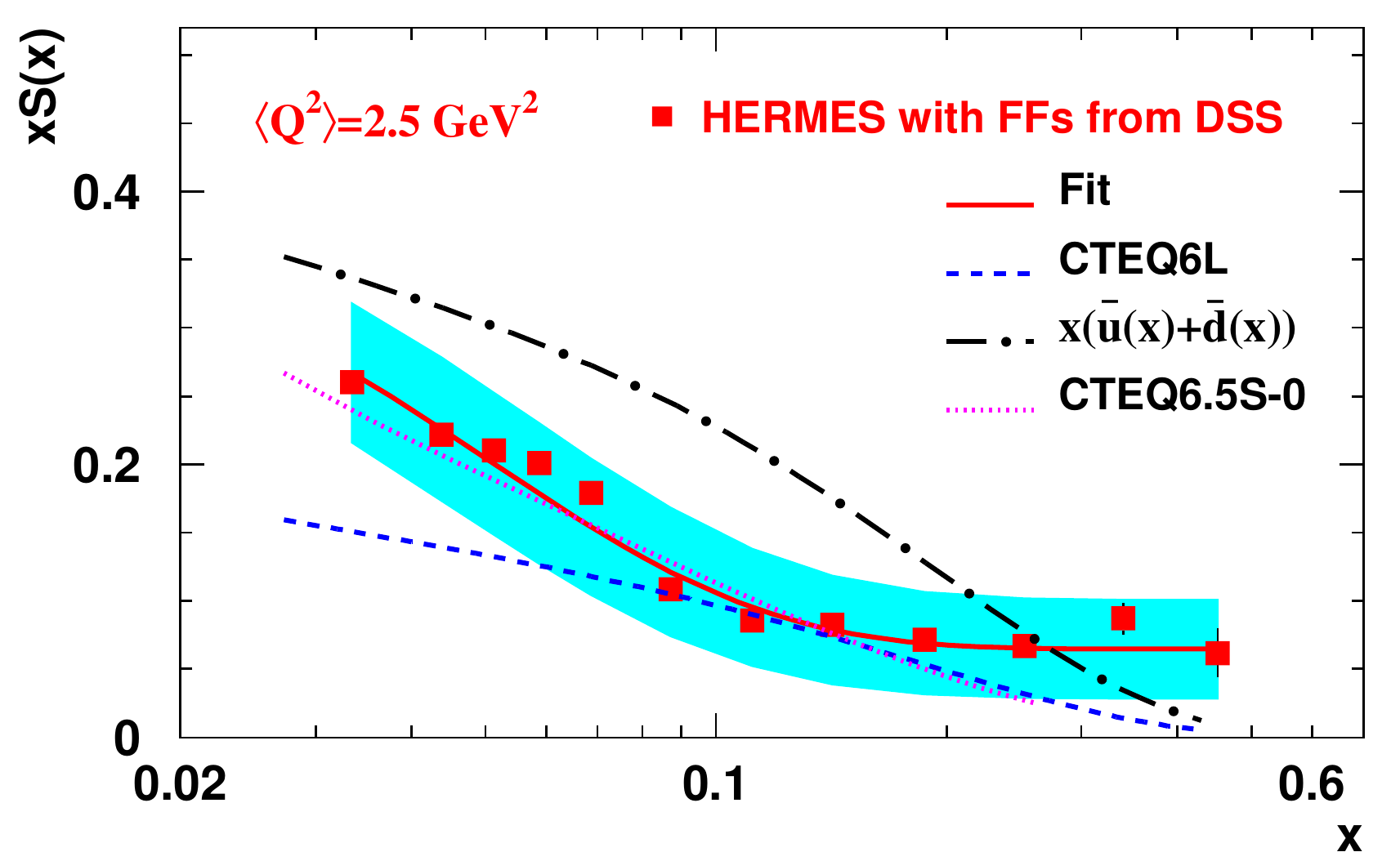}
	\caption{\label{fig:dss} 
	The strange-parton distribution $xS(x,{\mathrm{Q}}^2)$ from
	the measured HERMES multiplicity for charged kaons evolved to
	${\mathrm{Q}}^2=2.5$ GeV$^2$ using the compilation of 
	fragmentation functions (FFs) from~\cite{deFlorian:2007aj} (DSS). 
	The solid curve is a three-parameter
	fit to the data, the dashed curve
	gives $xS(x)$ from {CTEQ6L}, and the dot-dash curve is the sum of light
	antiquarks from {CTEQ6L}.
	The dotted curve is from {CTEQ6.5S-0}, a PDF reference 
	set~\cite{Lai:2007dq}
	in which the shape of $xS(x)$ has not been constrained.
	The band containing the experimental points represents the
	fully correlated systematic uncertainties arising from the 
	imprecision of ${\int{\cal D}^{K}_{Q}(z,{\mathrm{Q}}^2)dz}$.}
\end{figure}

To assess the impact of the more accurate estimate of 
${\int{\cal D}^{K}_{Q}(z,{\mathrm{Q}}^2)dz}$ resulting from the
HERMES data an extraction of $S(x,Q^2)$ was made using the
fragmentation functions from the global analysis of~\cite{deFlorian:2007aj}.
The results, presented in Fig.~\ref{fig:dss},
have an $xS(x,Q^2)$ with large uncertainties
in the high-$x$ region, and demonstrate the
importance of a more precise value for ${\int{\cal D}^{K}_{Q}(z,{\mathrm{Q}}^2)dz}$
as is that extracted from the analysis reported here.

The extracted quantity $\int_{0.2}^{0.8}{\cal D}^K_Q(z,\mathrm{Q}^2)dz$
was used together
with values of $Q(x,{\mathrm{Q}}^2)$ from {CTEQ6L}~\cite{Pumplin:2002vw} 
and the measured multiplicity to obtain 
the product $S(x,{\mathrm{Q}}^2){\int{\cal D}^{K}_{S}(z,{\mathrm{Q}}^2)dz}$. 
As in Ref.~\cite{Airapetian:2008qf}, the contribution of the strange
quarks to the denominator in Eq.~\ref{eq:mult} 
was initially neglected  leading to 
\begin{eqnarray}\label{eq:scomp}
\lefteqn{S(x,{\mathrm{Q}}^2){\int{\cal D}^{K}_{S}(z,{\mathrm{Q}}^2)dz}\simeq } \\ 
& & Q(x,\mathrm{Q}^2)\left[5\frac{dN^{K}(x,\mathrm{Q}^2)}{dN^{DIS}(x,\mathrm{Q}^2)}-
{\int{\cal D}^{K}_{Q}(z,{\mathrm{Q}}^2)dz}\right]\, . \nonumber 
\end{eqnarray}
A small iterative correction was then made to account for the neglect of this term.
 
 \begin{figure}[b*]
	\centering
	\includegraphics[width=0.89\linewidth]{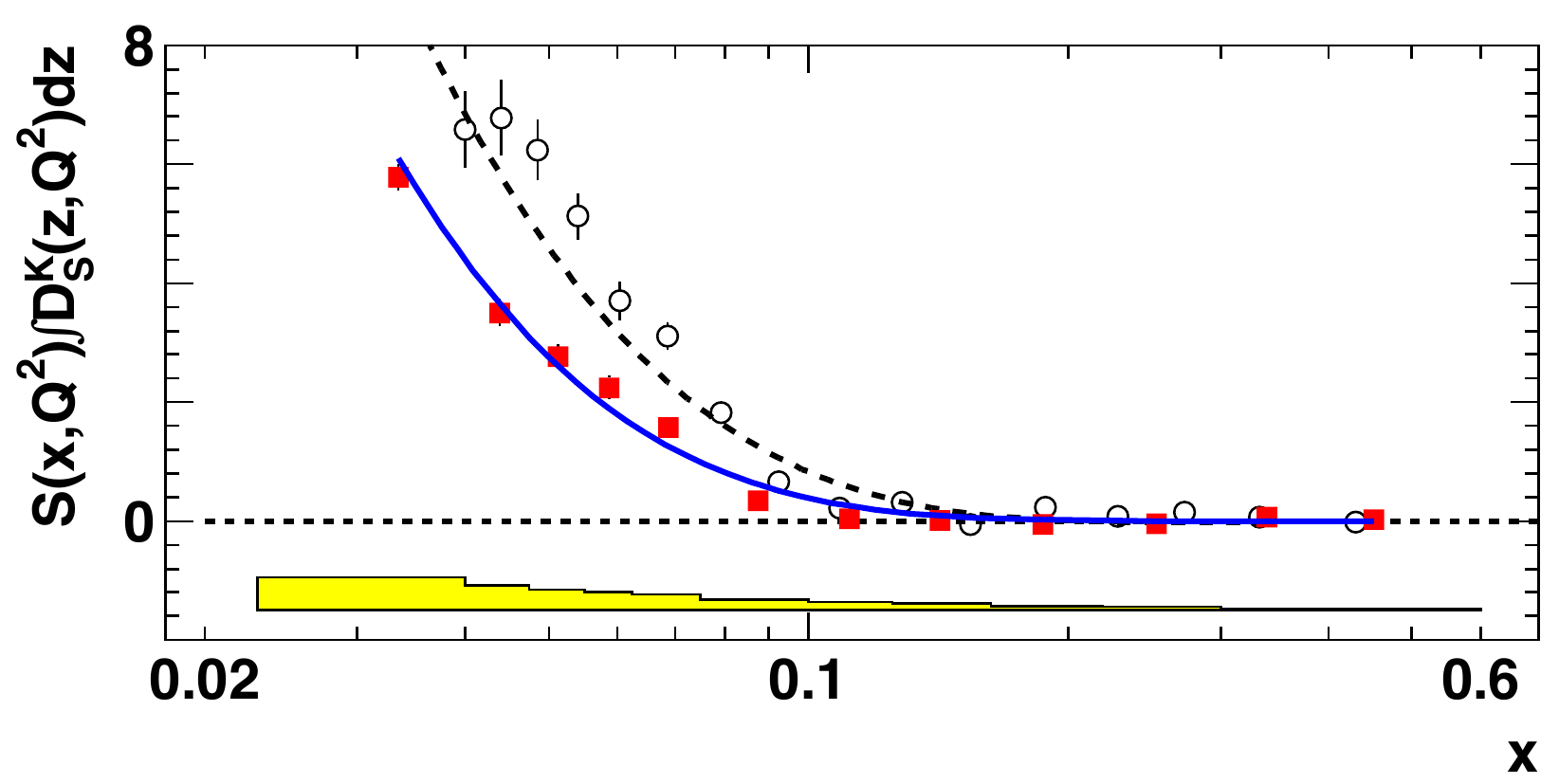}
	\caption{\label{fig:sxdz} The product,
	$S(x,{\mathrm{Q}}^2){\int{\cal D}^{K}_{S}(z,{\mathrm{Q}}^2)dz}$, of 
	the strange-quark PDF and the integral of the fragmentation
	function for strange quarks (squares) obtained from the measured
	HERMES multiplicity for charged kaons at the $\langle {\mathrm{Q}}^2 \rangle$
	for each bin. The solid curve is a least-squares
	fit with the result $f(x)=x^{-0.834\pm 0.019 }e^{-x/(0.0337\pm 0.0014)}(1-x)$. 
	The band represents propagated experimental systematic
	uncertainties. The open points and the dashed curve show the data
	and fit published previously in Ref.~\cite{Airapetian:2008qf}.}
\end{figure}

The result for the product 
$S(x,{\mathrm{Q}}^2){\int{\cal D}^{K}_{S}(z,{\mathrm{Q}}^2)dz}$
together with a fit 
of the form $x^{-a_1}e^{-x/a_2}(1-x)$ are shown in Fig.~\ref{fig:sxdz}.
In the region $x<0.1$ the values of the product are substantially
smaller than those reported previously~\cite{Airapetian:2008qf}.
This fit leads to the solid curve shown in Fig.~\ref{fig:mult_x}.
The use of the most recent NNPDF2.3LO reference PDF 
set~\cite{Rojo:2013hj} in place of
the {CTEQ6L} PDFs does not
alter significantly the results of the extraction.

\begin{figure}[t]
	\centering
	\includegraphics[width=.88\linewidth]{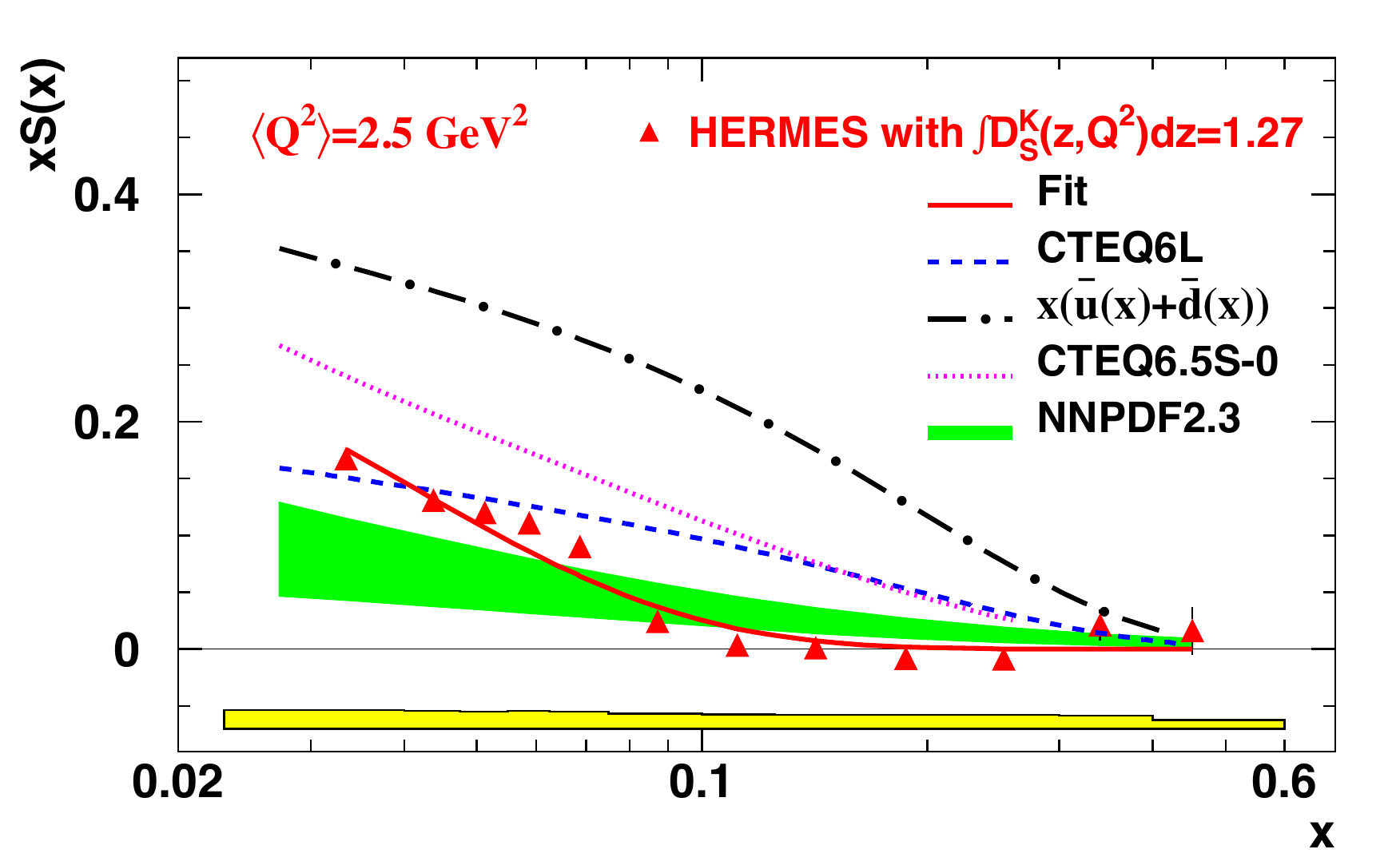}
	\caption{\label{fig:sxdz2p5} 
	The strange-parton distribution $xS(x,{\mathrm{Q}}^2)$ from
	the measured HERMES multiplicity for charged kaons evolved to
	${\mathrm{Q}}^2=2.5$ GeV$^2$ assuming ${\int{\cal D}^{K}_{S}(z,{\mathrm{Q}}^2)dz}=1.27$. 
	The solid curve is a two-parameter
	fit with $S(x)=[\int \mathcal{D}_S^K(z,{\mathrm{Q}}^2)dz]^{-1}\times
	x^{-0.867\pm 0.019}e^{-x/(0.0331\pm 0.0014)}(1-x)$. The dashed, dot-dash, dotted 
	curves are as given in Fig.~\ref{fig:dss}. 
	The broad band is the $\pm 1\sigma$
	zone of allowed values predicted by the neural network (NNPDF2.3) 
	reference set~\cite{Ball:2011uy}.
	The band at the bottom represents the propagated
	experimental systematic uncertainties.
	A scale uncertainty of approximately 10\%
	coming from the precision of ${\int{\cal D}^{K}_{S}(z,{\mathrm{Q}}^2)dz}$ is not shown.}
\end{figure}

In order to compare the distribution of $S(x,{\mathrm{Q}}^2)$ with the average of
those of the nonstrange quarks, the HERMES result for
$S(x,{\mathrm{Q}}^2){\int{\cal D}^{K}_{S}(z,{\mathrm{Q}}^2)dz}$ has been evolved to 
${\mathrm{Q}}^2 = 2.5$ GeV$^2$. The ${\mathrm{Q}}^2$ evolution factors are taken 
from {CTEQ6L} and from the fragmentation function compilation given 
in Ref.~\cite{deFlorian:2007aj}. 
Corrections to the evolution due to higher-twist 
contributions are assumed to be negligible, because higher-twist effects are
expected to be significant only for larger values
of $x$~\cite{Martin:1998np}, where the 
extracted distribution of $xS(x,{\mathrm{Q}}^2)$ vanishes.
The distribution of $xS(x,{\mathrm{Q}}^2)$ was 
obtained from $S(x,{\mathrm{Q}}^2){\int{\cal D}^{K}_{S}(z,{\mathrm{Q}}^2)dz}$ 
by dividing it by
${\int{\cal D}^{K}_{S}(z,{\mathrm{Q}}^2)dz}=1.27$, 
the value at ${\mathrm{Q}}^2=2.5$ GeV$^2$ given  
in~\cite{deFlorian:2007aj}.   
The uncertainty on ${\int{\cal D}^{K}_{S}(z,{\mathrm{Q}}^2)dz}$ enters only as a scale uncertainty in the extracted $xS(x,{\mathrm{Q}}^2)$.
The results are presented in Fig.~\ref{fig:sxdz2p5}.

Due to the anti-correlation of strange and non-strange kaon 
fragmentation functions in a global analysis, a proper consideration of 
the non-strange kaon fragmentation function obtained here 
may lead to a considerably smaller strange kaon fragmentation function.
Such a revision can be expected in the next global analysis, with
the result that the strange distribution as extracted here may increase.

As in the earlier extraction, the normalization of the HERMES points is
determined by the value of ${\int{\cal D}^{K}_{S}(z,{\mathrm{Q}}^2)dz}$ assumed. 
The values of the extracted distribution of $S(x,{\mathrm{Q}}^2)$ are 
smaller than those
reported in Ref.~\cite{Airapetian:2008qf}. But still, the
qualitative features of the shape of $xS(x,{\mathrm{Q}}^2)$ are strikingly 
different from
the shape of $xS(x,{\mathrm{Q}}^2)$ obtained with {CTEQ6L}  and 
other global QCD
fits of LO PDFs as well as that
of the sum of the light antiquarks. The absence of 
strength above $x\approx 0.1$ is clearly discrepant with
{CTEQ6L}. 
While, in principle, the new values
for the kaon multiplicities and fragmentation integrals reported 
here could significantly alter the results of 
the strange-quark helicity-distribution extraction
reported in Ref.~\cite{Airapetian:2008qf}, in fact, their use produces no
significant change in the helicity distribution reported there.

In conclusion, a new extraction of the multiplicities for charged kaons in 
DIS has been made and the extraction of the distribution of strange 
quarks in the nucleon has been reevaluated using these new data. 
In the measured range of $x$, the strength of the
polarization-averaged PDF $S(x,{\mathrm{Q}}^2)$ is,
under the same assumptions, substantially less than reported 
in~\cite{Airapetian:2008qf}, 
but the shape is similar, and the momentum
density is softer than that determined from the analysis of other experiments. 

\section*{Acknowledgements}

We warmly thank Juan Rojo for his efforts in generating an unpublished
NNPDF2.3LO PDF data set that includes the kinematic region of the HERMES experiment, 
and we gratefully acknowledge the DESY management for its support, the staff
at DESY and the collaborating institutions for their significant effort,
and our national funding agencies for financial support.

\bibliography{dc94}

\end{document}